\documentstyle[prl,aps,preprint,tighten]{revtex}

\begin{document}
\preprint{cond-mat/9504046}
\draft
\title{Insensitivity to time-reversal symmetry breaking of \\ universal
conductance fluctuations with Andreev reflection}
\author{P. W. Brouwer and C. W. J. Beenakker}
\address{Instituut-Lorentz, University of Leiden, P.O. Box 9506, 2300 RA
Leiden, The Netherlands}
\date{Received 10 February 1995}
\maketitle%

\begin{abstract}
Numerical simulations of conduction through a disordered microbridge between a
normal metal and a superconductor have revealed an anomalous insensitivity of
the conductance fluctuations to a magnetic field. A theory for the anomaly is
presented: Both an exact analytical calculation (using random-matrix theory)
and a qualitative symmetry argument (involving the exchange of time-reversal
for reflection symmetry).
\medskip
%
\end{abstract}


Universal conductance fluctuations (UCF) are a fundamental manifestation of
phase-coherent transport in disordered metals \cite{Altshuler,LeeStone}. The
adjective ``universal'' describes two aspects of the sample-to-sample
fluctuations of the conductance: (1) The variance $\mbox{var}\, G$ is of order
$(e^2/h)^2$, independent of sample size or disorder strength; (2) $\mbox{var}\,
G$ decreases precisely by a factor of two if time-reversal symmetry ({${\cal
T}$}) is broken by a magnetic field. The universality of this factor of two has
been established both by diagrammatic perturbation theory
\cite{Altshuler,LeeStone} and by random-matrix theory
\cite{Imry,Muttalib,Mello1988,StoneReview}. In the former approach, one has two
classes of diagrams, cooperons and diffusons, which contribute equally to
$\mbox{var}\, G$ in the presence of {${\cal T}$}. A magnetic field suppresses
the cooperons but leaves the diffusons unaffected, hence $\mbox{var}\, G$ is
reduced by $\case{1}{2}$. In the latter approach, the universality of the
factor-of-two reduction follows from the Dyson-Mehta theorem \cite{DysonMehta},
which applies to the variance $\mbox{var}\, A$ of any observable $A = \sum_n
a({T}_n)$ which is a linear statistic on the transmission eigenvalues ${T}_n$
\cite{Beenakker1993}. The crossover from a linear to a quadratic eigenvalue
repulsion upon breaking {${\cal T}$} directly leads to a reduction by
$\case{1}{2}$ of $\mbox{var}\, A$\cite{noot1}.

The situation is qualitatively different if the normal-metal conductor (N) is
attached at one end to a superconductor (S). At the NS interface the
dissipative normal current is converted into a dissipationless supercurrent,
via the scattering process of Andreev reflection \cite{Andreev}: An electron
incident from N is reflected as a hole, with the addition of a Cooper pair to
the superconducting condensate. The conversion from normal to supercurrent has
essentially no effect on the average conductance, provided that the interface
resistance is negligibly small \cite{BeenakkerLesHouches}. However, the effect
on the conductance fluctuations is striking: The variance is still universally
of order $(e^2/h)^2$, but it has become {\em insensitive} to the breaking of
{${\cal T}$}. Numerical simulations by Marmorkos, Jalabert, and one of the
authors \cite{Marmorkos} of a disordered wire attached to a superconductor have
shown that the variance is unaffected by a {${\cal T}$}-breaking magnetic
field, within the $10$\% statistical uncertainty of the simulations. This does
not contradict the Dyson-Mehta theorem, because the conductance $G_{\rm NS}$ of
the NS junction is a linear statistic in the presence --- but not in the
absence of {${\cal T}$} \cite{Beenakker1992}. One wonders whether there is some
hidden symmetry principle which would constrain $\mbox{var}\, G_{\rm NS}$ to be
the same, regardless of whether {${\cal T}$} is broken or not. No such symmetry
principle has been found, and in fact we do not know of any successful
generalization  so far of the theory of UCF to quantities which are not linear
statistics \cite{noot2}.

Here we wish to announce that we have succeeded in the analytical calculation
of $\mbox{var}\, G_{\rm NS}$ in the absence of {${\cal T}$}, using techniques
from random-matrix theory. We find that $\mbox{var}\, G_{\rm NS}$ for a
disordered wire attached to a superconductor is reduced by $(2-90/\pi^4)^{-1}
\approx 0.929$ upon breaking {${\cal T}$}. This number is sufficiently close to
$1$ to be consistent with the numerical simulations \cite{Marmorkos}, and
sufficiently different from $1$ to explain why attempts to find a rigorous
symmetry principle had failed. Still, we have been able to find an approximate
symmetry argument, which explains in an intuitively appealing way why the
number $(2-90/\pi^4)^{-1}$ is close to $1$. Our theory is more generally
applicable than to a disordered wire: It applies to any NS junction for which
the probability distribution $P(S)$ of the scattering matrix $S$ of the normal
region depends only on the transmission eigenvalues $T_n$. (Such a distribution
is called ``isotropic'' \cite{StoneReview}.) As two examples we consider a
disordered metal grain and a ballistic constriction in a disordered wire.

Starting point of our calculation is the general relation between the
conductance of the NS junction and the scattering matrix $S$ of the normal
region \cite{Beenakker1992},
\begin{mathletters} \label{She}
\begin{eqnarray}
  && G_{\rm NS} = 2 G_0\, \mbox{tr}\, m m^{\dagger}, \ \ G_0 \equiv 2 e^2 /h,
\\
  && m = \sqrt{{T}} (1 + {u} \sqrt{R} {u}^* \sqrt{R})^{-1} {u} \sqrt{{T}},\ \
{u} \equiv {w}_2^{\vphantom{*}} {w}_1^*.
\end{eqnarray}
\end{mathletters}%
We used the polar decomposition
\begin{equation}
  S = \left(\begin{array}{ll} {v}_1 & 0 \\ 0 & {w}_1 \end{array} \right)
      \left(\begin{array}{ll} {\rm i} \sqrt{R} & \sqrt{{T}} \\ \sqrt{{T}} &
{\rm i} \sqrt{R} \end{array} \right)
      \left(\begin{array}{ll} {v}_2 & 0 \\ 0 & {w}_2 \end{array} \right),
      \label{polar}
\end{equation}
where ${v}_1$, ${v}_2$, ${w}_1$, and ${w}_2$ are $N \times N$ unitary matrices
($N$ being the number of propagating modes at the Fermi level in each of the
two leads attached to the sample). The matrix ${T}$ is a diagonal matrix with
the $N$ transmission eigenvalues ${T}_i \in [0,1]$ on the diagonal, and $R = 1
- {T}$. In the presence of {${\cal T}$}, one has $S = S^{\rm T}$, hence ${w}_2
= {w}_1^{\rm T}$, hence ${u}=1$. (The superscript ${\rm T}$ denotes the
transpose of the matrix.) Eq.\ (\ref{She}) then simplifies to
\cite{Beenakker1992}
\begin{equation}
  G_{\rm NS}(\mbox{{${\cal T}$}}) = 2 G_0 {\textstyle \sum_{n}} {T}_n^2 (2 -
{T}_n)^{-2},
\end{equation}
and $\mbox{var}\, G_{\rm NS}$ follows directly from general formulas for the
variance of a linear statistic on the transmission eigenvalues
\cite{Beenakker1993,BeenakkerRChalkerM}. In the absence of {${\cal T}$} no such
simplification occurs.

To compute $\mbox{var}\, G_{\rm NS} = \langle G_{\rm NS}^2 \rangle - \langle
G_{\rm NS} \rangle^2$ in the absence of {${\cal T}$}, we assume an isotropic
distribution of $S$, which implies that the average $\langle \cdots \rangle$
over the ensemble of scattering matrices can be performed in two steps:
$\langle \cdots \rangle = \langle \langle \cdots \rangle_{{u}} \rangle_T$,
where $\langle\cdots\rangle_{{u}}$ and $\langle\cdots\rangle_T$ are,
respectively, the average over the unitary matrices ${{u}}$ and over the
transmission eigenvalues ${T}_i$. It is convenient to add and subtract $\langle
\langle G_{\rm NS} \rangle^2_{{u}} \rangle_T^{\vphantom{2}}$, so that the
variance of the conductance splits up into two parts,
\begin{eqnarray}
  \mbox{var}\, G_{\rm NS} &=& \left\langle \langle G_{\rm NS}^{\vphantom{2}}
\rangle^2_{{u}} \right\rangle_T^{\vphantom{2}} - \left\langle \langle G_{\rm
NS}^{\vphantom{2}} \rangle_{{u}}^{\vphantom{2}} \right\rangle_T^2 + \nonumber
\\ && \left\langle \langle G_{\rm NS}^2 \rangle_{{u}}^{\vphantom{2}} - \langle
G_{\rm NS}^{\vphantom{2}} \rangle^2_{{u}} \right\rangle_T^{\vphantom{2}},
\label{VarGNSsplit}
\end{eqnarray}
which we evaluate separately.

The first part is the variance of $\langle G_{\rm NS}\rangle_{{u}}$ over the
distribution of transmission eigenvalues. As a consequence of the isotropy
assumption, the matrix ${{u}}$ is uniformly distributed in the group ${\cal
U}(N)$ of $N \times N$ unitary matrices \cite{StoneReview}. To evaluate
$\langle G_{\rm NS}\rangle_{{u}}$ we need to perform an integral over ${\cal
U}(N)$ of a rational function of ${{u}}$, according to Eq.\ (\ref{She}). Such
matrix integrals are notoriously difficult to evaluate in closed form
\cite{Hua}, but fortunately we only need the large-$N$ limit. Creutz
\cite{Creutz} and Mello \cite{Mello} have given general rules for the integral
over ${\cal U}(N)$ of polynomial functions of ${{u}}$. By applying these rules
we find that
\begin{eqnarray}
 && \langle \mbox{tr}\, {T} ({u} \sqrt{R} {u}^* \sqrt{R})^p {u} {T}
{u}^{\dagger} (\sqrt{R} {u}^{\rm T} \sqrt{R} {u}^{\dagger})^q \rangle_{{u}} =
\nonumber \\ && \ \ \ \delta_{pq} N \tau_1^2 (1 - \tau_1)^{2p} + {\cal O}(1),
\end{eqnarray}
where we have defined the trace $\tau_k = N^{-1} \sum_i {T}_i^k$. It follows
that, up to corrections of order unity,
\begin{equation}
  \langle G_{\rm NS} \rangle_{{u}} = 2 G_0 N \sum_{p = 0}^{\infty} \tau_1^2 (1
- \tau_1)^{2p} = {2 G_0 N \tau_1 \over 2 - \tau_1}. \label{GNSavgw}
\end{equation}
Since $\tau_k$ is a linear statistic, we know that its fluctuations are an
order $1/N$ smaller than the average \cite{StoneReview}. This implies that, to
leading order in $1/N$, $\mbox{var}\, f(\tau_k) = [f'(\tau_k)]^2 \,
\mbox{var}\, \tau_k$. The variance of Eq.\ (\ref{GNSavgw}) is therefore
\begin{eqnarray}
  \left\langle \langle G_{\rm NS} \rangle^2_{{u}}
\right\rangle_T^{\vphantom{2}} - \left\langle \langle G_{\rm NS}^{\vphantom{2}}
\rangle_{{u}}^{\vphantom{2}} \right\rangle_T^2 &=& {16 G_0^2 N^2 \mbox{var}\,
\tau_1 \over (2 - \langle \tau_1 \rangle)^4} + {\cal O}(1/N). \label{VarGNS1}
\end{eqnarray}
Note that the leading term in Eq.\ (\ref{VarGNS1}) is ${\cal O}(1)$.

We now turn to the second part of Eq.\ (\ref{VarGNSsplit}), which involves the
variance $\langle G_{\rm NS}^2 \rangle_{{u}}^{\vphantom{2}} - \langle G_{\rm
NS}^{\vphantom{2}} \rangle^2_{{u}}$ of $G_{\rm NS}$ over ${\cal U}(N)$ at fixed
transmission eigenvalues and subsequently an average over the ${T}_i$'s. The
calculation is similar in principle to that described in the preceding
paragraph, but many more terms contribute to leading order in $1/N$. Here we
only give the result,
\begin{eqnarray}
    \left\langle \langle G_{\rm NS}^2 \rangle_{{u}}^{\vphantom{2}} - \langle
G_{\rm NS}^{\vphantom{2}} \rangle^2_{{u}} \right\rangle_T &=& 16 G_0^2 \left(2
- \langle \tau_1 \rangle \right)^{-6} \langle \tau_1 \rangle^{-2}
    \left\{ 4 \langle \tau_1 \rangle^2 - 8 \langle \tau_1 \rangle^3 + 9 \langle
\tau_1 \rangle^4 - 4 \langle \tau_1 \rangle^5 + 2 \langle \tau_1 \rangle^6 -
\right. \nonumber \\ && \left. 4 \langle \tau_1 \rangle \langle \tau_2 \rangle
+ 2 \langle \tau_1 \rangle^2 \langle \tau_2 \rangle -  2 \langle \tau_1
\rangle^3 \langle \tau_2 \rangle - 2 \langle \tau_1 \rangle^4 \langle \tau_2
\rangle + 6 \langle \tau_2 \rangle^2 - 6 \langle \tau_1 \rangle \langle \tau_2
\rangle^2 + \right. \nonumber \\ && \left. 3 \langle \tau_1 \rangle^2 \langle
\tau_2 \rangle^2 - 4 \langle \tau_1 \rangle \langle \tau_3 \rangle + 6 \langle
\tau_1 \rangle^2 \langle \tau_3 \rangle - 2 \langle \tau_1 \rangle^3 \langle
\tau_3 \rangle \right\} + {\cal O}(1/N). \label{VarGNS2}
\end{eqnarray}
The sum of Eqs.\ (\ref{VarGNS1}) and (\ref{VarGNS2}) equals $\mbox{var}\,
G_{\rm NS}$, according to Eq.\ (\ref{VarGNSsplit}). The resulting expression
contains only moments of the transmission eigenvalues. This solves the problem
of the computation of $\mbox{var}\, G_{\rm NS}$ in the absence of {${\cal T}$},
since these moments are known.

For the application to a disordered wire (length $L$, mean free path $\ell$)
one has the variance \cite{LeeStone,Mello1988} $N^2 \mbox{var}\, \tau_1 =
\case{1}{15}$, and averages \cite{BeenakkerBuettiker} $\langle \tau_k \rangle =
\case{1}{2} (\ell /L) \Gamma(\case{1}{2}) \Gamma(k)/\Gamma(k+\case{1}{2})$.
Substitution into Eqs.\ (\ref{VarGNS1}) and (\ref{VarGNS2}) yields (in the
diffusive limit $\ell/L \rightarrow 0$)
\begin{equation}
  \mbox{var}\, G_{\rm NS}(\mbox{no {${\cal T}$}}) = \case{8}{15} G_0^2 \approx
0.533\, G_0^2. \label{VarWire2}
\end{equation}
This is to be compared with the known result in the presence of {${\cal T}$}
\cite{BeenakkerRChalkerM}
\begin{equation}
  \mbox{var}\, G_{\rm NS}(\mbox{{${\cal T}$}}) = (\case{16}{15} - 48\pi^{-4})
G_0^2 \approx 0.574\, G_0^2. \label{VarWire1}
\end{equation}
Breaking {${\cal T}$} reduces the variance by less than $10$\%, as advertised.

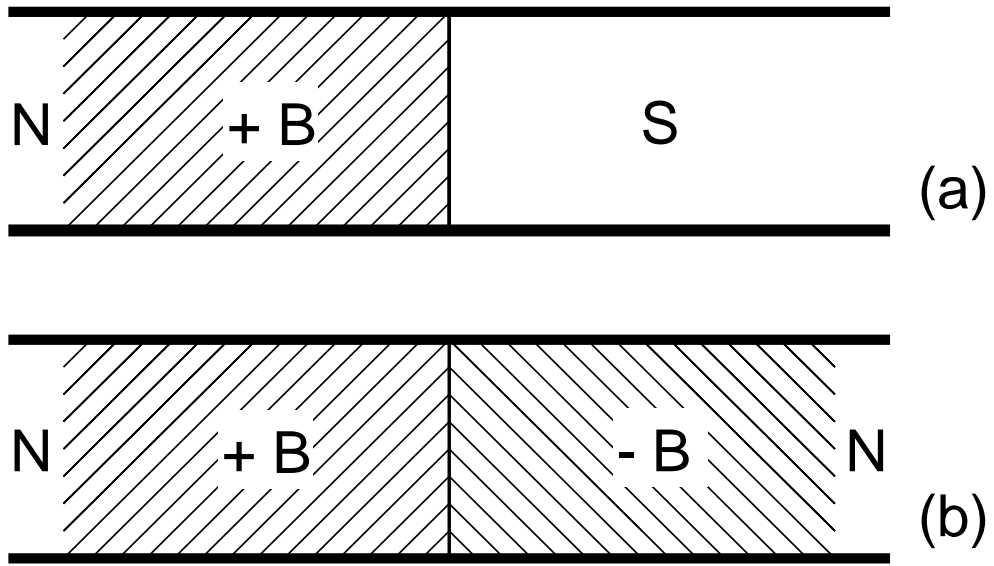
\begin{figure}[h]
\caption{\label{fig1} (a): Schematic drawing of a disordered normal metal (N)
connected to a superconductor (S), in a time-reversal symmetry ({${\cal T}$})
breaking magnetic field $B$. In (b) the normal region is connected in series
with its mirror image. As indicated, the magnetic field $B$ changes sign upon
reflection. The variance of the conductance fluctuations in (a) is exactly four
times the variance in (b). The variance in (b) is exactly two times the
variance in the absence of the reflection symmetry ({${\cal S}$}). The exchange
of {${\cal T}$} for {${\cal S}$} explains the insensitivity of the conductance
fluctuations to a magnetic field, as discussed in the text.}
\end{figure}

We would like to obtain a more direct understanding of why the two numbers in
Eqs.\ (\ref{VarWire2}) and (\ref{VarWire1}) are so close. To that end we return
to the general expression (\ref{She}) for the conductance $G_{\rm NS}$ of an NS
junction, in terms of the scattering matrix $S$ of the normal region. We
compare $G_{\rm NS}$ with the conductance $G_{\rm NN}$ of an entirely normal
metal consisting of two segments in series (see Fig.\ \ref{fig1}). The first
segment has scattering matrix $S$, the second segment is the mirror image of
the first. That is to say, the disorder potential is specularly reflected and
the sign of the magnetic field is reversed. The system NN thus has a reflection
symmetry ({${\cal S}$}), both in the presence and absence of {${\cal T}$}. The
scattering matrix of the second segment is $\Sigma S \Sigma$, where $\Sigma$ is
a $2N \times 2N$ matrix with zero elements, except for $\Sigma_{i,N+i} =
\Sigma_{N+i,i} = 1$ ($i=1,2,\ldots,N$). (The matrix $\Sigma$ interchanges
scattering states incident from left and right.) The conductance $G_{\rm NN}$
follows from the transmission matrix through the two segments in series by
means of the Landauer formula,
\begin{mathletters}\label{SNN}
\begin{eqnarray}
&& G_{\rm NN}(\mbox{{${\cal S}$}}) = G_0\, \mbox{tr}\, m' m'^{\dagger}, \\
&&  m'      = \sqrt{{T}} (1 + {{u}}' \sqrt{R} {{u}}' \sqrt{R})^{-1} {{u}}'
\sqrt{{T}}, \ \ {{u}}' \equiv {w}_2 {w}_1.
\end{eqnarray}
\end{mathletters}%
The difference between Eqs.\ (\ref{She}) and (\ref{SNN}) is crucial in the
presence of {${\cal T}$}, when ${w}_2 = {w}_1^{\rm T}$, so that ${{u}} = 1$
while ${{u}}'$ is some random (symmetric) unitary matrix. However, in the
absence of {${\cal T}$}, ${w}_1$ and ${w}_2$ are independent, so that both
${{u}}$ and ${{u}}'$ are randomly distributed unitary matrices. We have
repeated the calculation of the variance starting from Eq.\ (\ref{SNN}), and
found that $\mbox{var}\, \mbox{tr}\, m m^{\dagger} = \mbox{var}\, \mbox{tr}\,
m' m'^{\dagger}$, hence
\begin{equation}
  \mbox{var}\, G_{\rm NS}(\mbox{no {${\cal T}$}}) = 4\, \mbox{var}\, G_{\rm NN}
(\mbox{{${\cal S}$}, no {${\cal T}$}}). \label{VarNSNN}
\end{equation}

The system NN is special because it possesses a reflection symmetry. Breaking
{${\cal S}$} amounts to the replacement of the mirror-imaged segment by a
different segment, with scattering matrix $S'$ which is independent of $S$ but
drawn from the same ensemble. Breaking {${\cal S}$} reduces the variance of the
conductance fluctuations by a factor of two, regardless of whether {${\cal T}$}
is present or not,
\begin{equation}
  \mbox{var}\, G_{\rm NN}(\mbox{{${\cal S}$}}) = 2\, \mbox{var}\, G_{\rm NN}
(\mbox{no {${\cal S}$}}). \label{VarNNNN}
\end{equation}
We have checked this relation by an explicit calculation, but it seems
intuitively obvious if one considers that the eigenstates separate into even
and odd states which fluctuate independently. Since breaking {${\cal T}$} by
itself reduces the variance of $G_{\rm NN}$ by a factor of two, we may write
\begin{equation}
  \mbox{var}\, G_{\rm NN}(\mbox{{${\cal S}$}, no {${\cal T}$}}) = \mbox{var}\,
G_{\rm NN} (\mbox{{${\cal T}$}, no {${\cal S}$}}). \label{VarTRSSRS}
\end{equation}
Eqs.\ (\ref{VarNSNN})--(\ref{VarTRSSRS}) are exact, and hold for any isotropic
distribution of the scattering matrix. We need one more relationship, which is
approximate and holds only for the case of a disordered wire
\cite{Beenakker1992,TakaneEbisawa}:
\begin{equation}
  \mbox{var}\, G_{\rm NS}(\mbox{{${\cal T}$}}) \approx 4 \mbox{var}\, G_{\rm
NN} (\mbox{{${\cal T}$}, no {${\cal S}$}}). \label{VarNSNNTRS}
\end{equation}
Taken together, Eqs.\ (\ref{VarNSNN})--(\ref{VarNSNNTRS}) imply the approximate
relationship $\mbox{var}\, G_{\rm NS}(\mbox{{${\cal T}$}}) \approx \mbox{var}\,
G_{\rm NS}(\mbox{no {${\cal T}$}})$. The exact calculation shows that the
approximation is accurate within $10$\%. We now understand the insensitivity of
the conductance fluctuations to a magnetic field as an {\em exchange of
symmetries}: Breaking {${\cal T}$} induces {${\cal S}$}, thereby compensating
the reduction of $\mbox{var}\, G_{\rm NS}$.

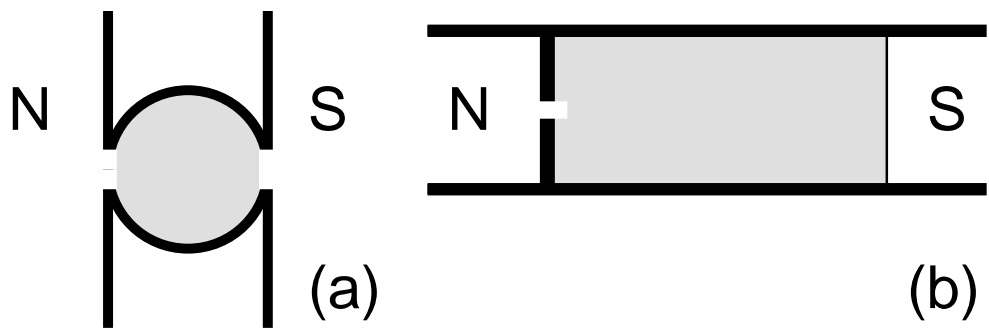
\begin{figure}[h]
\caption{\label{fig2} (a): Schematic drawing of an NS junction consisting of a
disordered metal grain (shaded). (b): A disordered normal-metal wire (shaded)
containing a point contact.}
\end{figure}

We have emphasized the general applicability of Eqs.\ (\ref{VarGNS1}),
(\ref{VarGNS2}) and (\ref{VarNSNN})--(\ref{VarTRSSRS}), which hold not just for
a disordered wire, but for any isotropic distribution of the scattering matrix.
We illustrate this by two examples. The first is an NS junction consisting of a
disordered metal grain (see Fig.\ \ref{fig2}a). The coupling of N and S to the
grain occurs via ballistic point contacts (width much smaller than the mean
free path in the grain). Following Ref.\ \ref{BarangerMJalabertPB}, we may
assume that the scattering matrix of the grain is distributed according to the
circular ensemble of random-matrix theory. This is an isotropic distribution.
The relevant moments of the transmission eigenvalues in the absence of {${\cal
T}$} are \cite{BarangerMJalabertPB} $\langle \tau_k \rangle = \pi^{-1/2}
\Gamma(k + \case{1}{2})/\Gamma(k+1)$, $N^2 \mbox{var}\, \tau_1 = \case{1}{16}$.
Substitution into the general formulas (\ref{VarGNS1}) and (\ref{VarGNS2})
yields
\begin{equation}
  \mbox{var}\, G_{\rm NS}(\mbox{no {${\cal T}$}}) = \case{128}{243} G_0^2
\approx 0.527 \, G_0^2,
\end{equation}
which is again close to the known result in the presence of {${\cal T}$}
\cite{BarangerMJalabertPB},
\begin{equation}
  \mbox{var}\, G_{\rm NS}(\mbox{{${\cal T}$}}) = \case{9}{16} G_0^2 \approx
0.563\, G_0^2.
\end{equation}

The second example is a ballistic constriction (point contact) in a wire which
is connected to a superconductor (see Fig.\ \ref{fig2}b). The point contact has
conductance $N_0 G_0$, which we assume to be much smaller than the conductance
$N \ell/L$ of the disordered wire by itself. As discussed in Ref.\
\ref{BeenakkerMelsen}, we may assume an isotropic distribution of the
scattering matrix of the combined system (point contact plus disordered wire).
The moments of the transmission eigenvalues are \cite{BeenakkerMelsen} $\langle
\tau_k \rangle = N_0/N, N^2 \mbox{var}\, \tau_1 = {\cal O}(N_0 L/N \ell)^2$.
Substitution into Eqs.\ (\ref{VarGNS1}) and (\ref{VarGNS2}) yields, in the
limit $N_0 L/N \ell \rightarrow 0$,
\begin{equation}
  \mbox{var}\, G_{\rm NS}(\mbox{no {${\cal T}$}}) = \case{1}{2} G_0^2.
\label{PointContact}
\end{equation}
In contrast, if {${\cal T}$} is not broken, the conductance fluctuations are
suppressed in this limit \cite{BeenakkerMelsen,MaslovBK}:
\begin{equation}
  \mbox{var}\, G_{\rm NS}(\mbox{{${\cal T}$}}) = {\cal O}(N_0 L/N \ell)^2 \ll
G_0^2.
\end{equation}
In this geometry a magnetic field greatly enhances the conductance
fluctuations. The reason that a disordered wire with a constriction behaves so
differently from an unconstricted wire, is that the relation (\ref{VarNSNNTRS})
does not hold in the presence of a constriction. However, the general
relationship (\ref{VarNSNN}) does hold, and indeed the result
(\ref{PointContact}) is four times the variance of a structure consisting of
two point contacts in series with a reflection symmetry.

In summary, we have solved the problem of universal conductance fluctuations in
normal-metal--superconductor junctions in a magnetic field, under the
assumption of an isotropic distribution of the scattering matrix of the normal
region. We find that breaking time-reversal symmetry introduces an approximate
reflection symmetry into the structure of the scattering matrix. This
reflection symmetry compensates the reduction of the conductance fluctuations
due to breaking of time-reversal symmetry, and explains the anomalous
insensitivity of the fluctuations in a magnetic field discovered in computer
simulations \cite{Marmorkos}. We  know of no other situation in physics where
application of a magnetic field creates rather than destroys a spatial
symmetry.

Discussions on this problem with A.\ Altland are gratefully acknowledged. This
work was supported by the Dutch Science Foundation NWO/FOM.

\begin{references}
\bibitem{Altshuler} B. L. Al'tshuler, Pis'ma Zh.\ Eksp.\ Teor.\ Fiz.\ {\bf 41},
530 (1985) [JETP Lett.\ {\bf 41}, 648 (1985)].
\bibitem{LeeStone} P. A. Lee and A. D. Stone, Phys.\ Rev.\ Lett.\ {\bf 55},
{\tt 1622} (1985).
\bibitem{Imry} Y. Imry, Europhys.\ Lett.\ {\bf 1}, 249 (1986).
\bibitem{Muttalib} K. A. Muttalib, J.-L. Pichard, and A. D. Stone, Phys.\ Rev.\
Lett.\ {\bf 59}, 2475 (1987).
\bibitem{Mello1988} P. A. Mello, Phys.\ Rev.\ Lett.\ {\bf 60}, 1089 (1988).
\bibitem{StoneReview} A. D. Stone, P. A. Mello, K. A. Muttalib, and J.-L.
Pichard, in {\em Mesoscopic Phenomena in Solids}, edited by B. L. Al'tshuler,
P. A. Lee, and R. A. Webb (North--Holland, Amsterdam, 1991).
\bibitem{DysonMehta} F. J. Dyson and M. L. Mehta, J.\ Math.\ Phys.\ {\bf 4},
701 (1963).
\bibitem{Beenakker1993} C. W. J. Beenakker, Phys.\ Rev.\ Lett.\ {\bf 70}, 1155
(1993).
\bibitem{noot1} Here, and in the rest of the paper, we assume that there is no
spin-orbit interaction, and that spin-rotation symmetry is maintained both with
and without {${\cal T}$}.
\bibitem{Andreev} A. F. Andreev, Zh.\ Eksp.\ Teor.\ Fiz.\ {\bf 46}, 1823 (1964)
[Sov.\ Phys.\ JETP {\bf 19}, 1228 (1964)].
\bibitem{BeenakkerLesHouches} For a review, see: C. W. J. Beenakker, in  {\em
Mesoscopic Quantum Physics}, edited by E. Akkermans, G. Montambaux, J.-L.
Pichard, and J. Zinn-Justin (North-Holland, Amsterdam, 1995).
\label{BeenakkerLesHouches}
\bibitem{Marmorkos} I. K.  Marmorkos, C. W. J.  Beenakker, and R. A.  Jalabert,
Phys.\ Rev.\ B {\bf 48}, 2811 (1993).\label{Marmorkos}
\bibitem{Beenakker1992} C. W. J. Beenakker, Phys.\ Rev.\ B {\bf 46}, 12841
(1992).
\bibitem{noot2} A promising field-theoretic approach to this problem, based on
the mapping onto a supersymmetric non-linear $\sigma$-model, has so far not
been successful [A. Altland, private communication]. The more conventional
diagrammatic perturbation theory suffers form a proliferation of relevant
diagrams, and has so far not been completed even in the presence of {${\cal
T}$} [Y. Takane and H. Ebisawa, J.\ Phys.\ Soc.\ Japan {\bf 60}, 3130 (1991)].
\bibitem{BeenakkerRChalkerM} C. W. J. Beenakker and B. Rejaei, Phys.\ Rev.\
Lett.\ {\bf 71}, 3689 (1993); J. T. Chalker and A. M. S. Mac\^edo, Phys.\ Rev.\
Lett.\ {\bf 71}, 3693 (1993).
\bibitem{Hua} L.\ K.\ Hua, {\em Harmonic Analysis of Functions of Several
Complex Variables in the Classical Domains} (Amer.\ Math.\ Soc.,\ Providence,
1963). \label{Hua}
\bibitem{Creutz} M.\ Creutz, J.\ Math.\ Phys.\ {\bf 19}, 2043 (1978).
\bibitem{Mello} P. A. Mello, J.\ Phys.\ A {\bf 23}, 4061 (1990).
\bibitem{BeenakkerBuettiker} C. W. J. Beenakker and M. B\"uttiker, Phys.\ Rev.\
B {\bf 46}, 1889 (1992).
\bibitem{TakaneEbisawa} Y. Takane and H. Ebisawa, J. Phys.\ Soc.\ Japan {\bf
61}, 2858 (1992).
\bibitem{BarangerMJalabertPB} \label{BarangerM} H. U. Baranger and P. A. Mello,
Phys.\ Rev.\ Lett.\ {\bf 73}, 142 (1994); R. A. Jalabert, J.-L. Pichard, and C.
W. J. Beenakker, Europhys.\ Lett.\ {\bf 27}, 255 (1994).
\label{BarangerMJalabertPB}
\bibitem{BeenakkerMelsen} C. W. J. Beenakker and J. A. Melsen, Phys.\ Rev.\ B
{\bf 50}, 2450 (1994). \label{BeenakkerMelsen}
\bibitem{MaslovBK} D. L. Maslov, C. Barnes, and G. Kirczenow, Phys.\ Rev.\
Lett.\ {\bf 70}, 1984 (1993).
\end{references}
\end{document}